\documentclass[final,5p,times,twocolumn,sort&compress]{elsarticle}
\usepackage{amssymb}
\usepackage{lipsum}
\usepackage{color} 
\usepackage{graphicx}
\usepackage{dcolumn}
\usepackage{bm}
\usepackage{physics}
\usepackage{hyperref}
\usepackage{slashed}
\def\u1{\textrm{U}(1)}
\def\su2u1{SU(2)$_L\times$U(1)$_Y$}
\newenvironment{Eqnarray}%
     {\arraycolsep 0.14em\begin{eqnarray}}{\end{eqnarray}}
\def\beq{\begin{equation}}
\def\eeq{\end{equation}}
\def\beqa{\begin{Eqnarray}}
\def\eeqa{\end{Eqnarray}}
\def\eq#1{Eq.~(\ref{#1})}
\def\Eq#1{Eq.~(\ref{#1})}
\def\Eqs#1#2{Eqs.~(\ref{#1}) and (\ref{#2})}
\def\eqs#1#2{Eqs.~(\ref{#1}) and (\ref{#2})}

\def\eqst#1#2{Eqs.~(\ref{#1})--(\ref{#2})}

\def\phm{\phantom{-}}
\def\lsub#1{_{\lower 1.5pt\hbox{$\scriptstyle#1$}}}
\def\lsim{\mathrel{\raise.3ex\hbox{$<$\kern-.75em\lower1ex\hbox{$\sim$}}}}
\def\gsim{\mathrel{\raise.3ex\hbox{$>$\kern-.75em\lower1ex\hbox{$\sim$}}}}

\journal{Physics Letters B}

\begin{document}

\begin{frontmatter}

\title{Classes of complete dark photon models constrained by 
\texorpdfstring{$Z$}{Z}-physics}

\author[first]{Miguel P. Bento}
\affiliation[first]{organization={CFTP, Departamento de Física, Instituto Superior Técnico,
Universidade de Lisboa},
            addressline={Avenida Rovisco Pais 1}, 
            postcode={Lisboa 1049}, 
            country={Portugal}}

\author[second]{Howard E. Haber}
\affiliation[second]{organization={Santa Cruz Institute for Particle Physics,
University of California},
            addressline={1156 High Street}, 
            city={Santa Cruz},
            postcode={CA 95064}, 
            country={USA}}

\author[first]{João P. Silva}

\begin{abstract}
Dark Matter models that employ a vector portal to a dark sector
are usually treated as an effective theory that incorporates kinetic mixing of the photon
with a new U(1) gauge boson, with the $Z$ boson integrated out.
However, a more complete theory must employ the full
SU(2)$_L\times $U(1)$_Y \times $U(1)$_{Y^\prime}$ gauge group, in which
kinetic mixing of the $Z$ boson with the new U(1) gauge boson
is taken into account.  The importance of the more complete analysis is 
demonstrated by an example where the parameter space 
of the effective theory that yields the observed dark matter relic density
is in conflict with a suitably defined electroweak $\rho$ parameter
that is deduced from a global fit to $Z$ physics data.
\end{abstract}


\end{frontmatter}

\section{\label{sec:intro}Introduction}

The Standard Model (SM) of electroweak interactions 
based on an SU(2)$_L\times$U(1)$_Y$ gauge theory describes to very
high precision the fundamental particles discovered in the laboratory~\cite{Grossman:2023wrq}.
Yet, it is known from cosmological and astrophysical observables, that
SM particles constitute only around 15\% of the total matter
content of the Universe, with the remaining 85\% consisting of non-baryonic dark matter (DM)
that cannot be accommodated by the SM~\cite{Planck:2018vyg,Profumo:2017hqp,Bauer:2017qwy}.
Some of the leading contenders for physics beyond the SM 
that contain candidates for particles that can contribute to the DM 
include models where the usual massless electromagnetic photon,
associated with the unbroken U(1)$_\textrm{EM}$ gauge group,
couples through kinetic mixing~\cite{Okun:1982xi,Galison:1983pa,Holdom:1985ag,Foot:1991kb,Babu:1997st}
with a very light massive dark photon~\cite{Essig:2010xa} (also called dark $Z'$ boson~\cite{Davoudiasl:2012ag,Alves:2013tqa})
that is associated with a new U(1)$_{Y'}$ gauge symmetry.  
The dark photon (or equivalently the dark $Z'$ boson) serves as a mediator between
the sector of SM particles and a dark sector which contains particles that
are neutral with respect to SM gauge group.   The DM can then be identified
with either stable particles and/or particles with lifetimes significantly
larger than the age of the Universe that reside in the dark sector (e.g., see \cite{Arkani-Hamed:2008kxc,Ackerman:2008kmp,Feng:2009mn,Chu:2011be,Davoudiasl:2012qa,Foot:2014uba,Curtin:2014cca,Foot:2014osa,An:2014twa,Alves:2015pea,Davoudiasl:2015bua,Backovic:2015fnp,Foot:2016wvj,Correia:2016xcs,Knapen:2017xzo,Darme:2017glc,Ilten:2018crw,Mondino:2020lsc,Foguel:2022unm,Harigaya:2023uhg,Davoudiasl:2023cnc}).
This analysis can and is usually made without referring to the $Z$ boson.
The idea is that the interactions relevant for DM studies
occur at an energy significantly below the $Z$ boson mass
($m_Z$), in which case this field can be integrated out.
One then proceeds to study the effective theory that depends solely
on two parameters.

However, a complete electroweak and DM model must
indeed start from the full SU(2)$_L\times $U(1)$_Y\times $U(1)$_{Y'}$ gauge theory.
The physical $Z$ boson field will then comprise a small component
of the U(1)$_{Y'}$ boson field, which affects the interpretation of the
precision electroweak observables measured at colliders.
Now the question arises:
in such complete theories, is there a tension between the
parameters required for a (low-energy) explanation for DM
and those required to conform with $Z$ boson observables?
A number of $Z$ physics observables have been 
considered in~\cite{Davoudiasl:2012ag,Davoudiasl:2012qa,
Curtin:2014cca,Davoudiasl:2015bua,Harigaya:2023uhg,Davoudiasl:2023cnc,Hook:2010tw,Sun:2023kfu}.
In this work, we examine tree-level corrections to the electroweak $\rho$ parameter, which is very
precisely determined in  \cite{EWPDG}.  This yields a very
important condition that the putative complete electroweak-DM model
must obey in order to comply with both dark matter searches and the collider constraints 
arising from precision electroweak studies.  

In Section~\ref{sec2}, we review the SU(2)$_L\times $U(1)$_Y\times $U(1)$_{Y'}$ 
gauge theory, which includes kinetic mixing between the U(1)$_Y$ and U(1)$_{Y'}$ 
gauge bosons, where the gauge bosons are
coupled to an arbitrary set of scalar multiplets.
Diagonalizing the neutral vector boson squared mass
matrix yields the photon, $Z$ boson, and the dark $Z'$ boson.
Following \cite{Bento:2023weq}, we define a 
suitable electroweak $\rho$ parameter and show that the SM tree-level result
of $\rho=1$ is modified.  In Section~\ref{sec:dark_matter}, 
a dark matter candidate is introduced by adding a Dirac fermion $\chi$
to the model
that is neutral with respect to the SM but has a nonzero U(1)$_{Y'}$ charge.  
Two different mass orderings for $m_{\chi}$ and $m_{Z'}$ are considered.
Conditions are then obtained on the model parameters such that $\chi$ constitutes
the observed dark matter.  In Section~\ref{sec4}, we provide an example where 
the latter result
is in conflict with the value of $\rho$ deduced from a global fit to $Z$ physics data.
In particular, as noted in our conclusions presented in Section~\ref{sec:Concl},
realistic models containing a dark $Z'$ as a vector
portal to a dark sector must be reconsidered in the framework of a gauge theory
that contains the $Z$ boson in addition to the photon and dark~$Z'$.  
Ultimately, one must check all constraints imposed by the precision electroweak data 
in order to achieve a consistent DM model.

\section{\label{sec:vector}An extra vector boson}
\label{sec2}

\subsection{Gauge sector}

The presence of an extra vector boson $\hat{X}$ allows for kinetic mixing
between abelian gauge bosons in the Lagrangian,

\begin{equation}
    \mathcal{L} \supset
    - \frac{1}{4} \hat{B}_{\mu \nu} \hat{B}^{\mu \nu}
    - \frac{1}{4} \hat{X}_{\mu \nu}  \hat{X}^{\mu \nu} 
    + \frac{\epsilon}{2 c_W} \hat{X}_{\mu \nu} \hat{B}^{\mu \nu} \, ,
\end{equation}
where $\hat{B}_\mu$ and $\hat{X}_\mu$ are the gauge bosons of $U(1)_Y$ and $U(1)_{Y'}$,
respectively, and $\hat{B}^{\mu \nu}$ and $\hat{X}^{\mu \nu}$ are the
corresponding field strength tensors.
The last term describes kinetic mixing 
between the U(1)$_Y$ and the U(1)$_{Y^\prime}$ gauge bosons, where
$c_W$ is defined implicitly in Eq.~\eqref{eq:def_sw} below.

Next, we transform the $\hat{B}$ and $\hat{X}$ fields
such that
\begin{equation}
    \hat{X}^\mu = \eta X^\mu  \, , \quad 
    \hat{B}^\mu = B^\mu + \frac{\epsilon}{c_W}  \eta X^\mu \, ,
\end{equation}
where 
\beq \label{etadef}
\eta\equiv\frac{1}{\sqrt{1 - \epsilon^2/c_W^2}}\,.
\eeq
We then recover the
canonical form of the kinetic Lagrangian, $\mathcal{L} \supset
    - \frac{1}{4} B_{\mu \nu} B^{\mu \nu}
    - \frac{1}{4} X_{\mu \nu}  X^{\mu \nu}$. 
Adding to this the kinetic Lagrangian of the SU(2)$_L$
fields ($W^\pm$ and $W^3$)
and defining the SU(2)$_L$, U(1)$_Y$ and U(1)$_{Y'}$ gauge couplings
by $g$, $g'$, and $g_X$, respectively, one can 
derive the mass eigenstates of the gauge bosons.

\subsection{Scalar multiplets}

All gauge bosons are massless before spontaneous gauge symmetry breaking.
In order to generate mass for the $Z$ boson and a very light dark photon
(henceforth denoted as the dark $Z'$),
one needs to break the gauge symmetry with some scalar fields.
Here, we consider a complex scalar doublet $\Phi$ with $Y'=0$ and SM quantum numbers 
(i.e., weak isospin $t_1=1/2$ and hypercharge $y_1=1/2$), and
$N-1$ additional real or complex scalar multiplets $\varphi_i$ (the latter charged under SU(2)$_L$, $Y$, and/or $Y'$
with corresponding weak isospin $t_i$ and U(1) charges $y_i$ and $y_i^\prime$, respectively, for $i=2,\ldots, N$),
each with an electrically neutral component.
The neutral components of the scalars acquire 
vacuum expectation values
 $\expval{\Phi^0} = v_1 / \sqrt{2}$
and $\expval{\varphi^0_i} = v_i / \sqrt{2} \, , \, i=2, \ldots N$,
which spontaneously break the gauge group. 
These, in turn, give mass
to the $W^\pm$ boson such that
\begin{equation}
    m_W^2 = \frac{g^2 v^2}{4} = \frac{g^2 \left[v_1^2 +
    \sum_{i=2}^N 2(C_{R_i} - y^2_i) v_i^2 c_i\right]}{4} \, ,
\end{equation}
where $C_{R_i} = t_i(t_i+1)$ for a complex [real] $\varphi_i$ multiplet,
with $c_i=1$ [$c_i=1/2$].
In order to approximately reproduce the observed value of $m_W/m_Z$, we shall
choose scalar field multiplets such that
\begin{equation}\label{eq:cr_3y2}
    C_{R_i} = 3 y_i^2 \,,
\end{equation}
which ensures that \eq{eq:rho_prime} is satisfied.

As for the neutral gauge bosons,
by mixing $B_\mu$ and $W^3_\mu$ in the usual way,
we identify the photon field $A_\mu$ (and the corresponding orthogonal field $Z_\mu^0$) by 
\begin{align}
    &A_\mu = W^3_\mu s_W + B_\mu c_W\,, \label{eq:WB_ZA1} \\
    &Z_\mu^0 = W^3_\mu c_W - B_\mu s_W \label{eq:WB_ZA2} \,,
\end{align}
where $c_W\equiv \cos\theta_W$ and $s_W\equiv\sin\theta_W$.
\eqs{eq:WB_ZA1}{eq:WB_ZA2} define the weak mixing angle $\theta_W$ such that $e = g s_W$ 
and $g' = g t_W$ (where $t_W\equiv s_W/c_W$). Indeed, we may define the weak mixing angle (at tree level) in terms
of physical observables via \cite{Bento:2023weq},
\begin{equation}\label{eq:def_sw}
    s_W^2 = \frac{\pi \alpha\lsub{\rm EM}}{\sqrt{2} G_F m_W^2} \,,
\end{equation}
where $G_F$ is the weak interaction Fermi constant and $\alpha_{\rm EM}\equiv e^2/(4\pi)$.
Note that Eq.~\eqref{eq:def_sw} 
can be applied
both in the SM and in U(1)$^\prime$ extended gauge theories.

The squared-mass matrix of the remaining (massive) neutral gauge bosons 
with respect to the $\{ Z^0, X \}$ basis is then given by
\begin{equation}\label{eq:mzzprime}
    \mathcal{M}^2
    =
    \begin{bmatrix}
       m_{Z^0}^2 & (\mathcal{M}^2)_{12} \\[10pt]
        (\mathcal{M}^2)_{12} &   (\mathcal{M}^2)_{22}
    \end{bmatrix} \, ,
\end{equation}
where $\eta$ is defined in \eq{etadef}, $\tau \equiv g_X / g$,
\begin{equation} \label{em12}
    (\mathcal{M}^2)_{12} = - \frac{m_{Z^0}^2}{v^2}\left[4 \eta t_W \epsilon \sum_{i=1}^N v_i^2 y_i^2
        +4 \eta \tau c_W \sum_{i=2}^N v_i^2 y_i y^\prime_i\right]\,,
\end{equation}
and 
\begin{equation}
    m_{Z^0} \equiv \frac{g v }{2 c_W} = \frac{m_W}{c_W} \, .
\label{eq:mZ_def}
\end{equation}
We are suppressing the explicit expression for $(\mathcal{M}^2)_{22}$, as it is not needed in what follows.
Note that the interaction eigenstate field $Z^0$ does \textit{not} correspond to the field of the experimentally observed $Z$ boson since it is \textit{not}
a mass eigenstate field, and the mass of the $Z$ (denoted below by $m_Z$) is \textit{not} equal to $m_{Z^0}$.

The mass eigenstate fields $Z$ and $Z'$ are obtained via
\begin{eqnarray}
&&
\left(
\begin{array}{cc}
Z^0 & X
\end{array}
\right)\ \mathcal{M}^2
\ \left(\begin{array}{c}
Z^0\\
X
\end{array}
\right) =
\nonumber\\[3pt]
&&
\hspace{7ex} \left(
\begin{array}{cc}
Z & Z^\prime
\end{array}
\right)\ 
\left(
\begin{array}{cc}
m_Z^2 & 0\\
0 & m_{Z^\prime}^2
\end{array}
\right)
\ \left(\begin{array}{c}
Z\\
Z^\prime
\end{array}
\right)\,,
\end{eqnarray}
where $m_{Z'}$ is the mass of the dark $Z'$ and
\begin{equation}
\left(\begin{array}{c}
Z^0\\
X
\end{array}
\right)
=
\left(
\begin{array}{cc}
\cos{\alpha} & - \sin{\alpha}\\
\sin{\alpha} & \phm\cos{\alpha}
\end{array}
\right)
\left(\begin{array}{c}
Z\\
Z^\prime
\end{array}
\right)
\end{equation}
defines the mixing angle $\alpha$.
We may then extract the important relation,
\beq  \label{eq:mz0_mz} 
m_{Z^0}^2 = m_Z^2 \cos^2 \alpha + m_{Z^\prime}^2 \sin^2 \alpha \,.
\eeq

In an SU(2)$_L\times $U(1)$_Y$ gauge theory, the choice of scalar multiplets that satisfy
Eq.~\eqref{eq:cr_3y2}
has been imposed in order to
ensure that the tree-level electroweak $\rho$ parameter,
\begin{equation} \label{treerho}
    \rho \equiv \frac{m_W^2}{m_Z^2 c_W^2},
\end{equation}
satisfies $\rho=1$
without resorting to a fine-tuning of the choice of scalar field vacuum expectation values.
Examples include multi-Higgs doublet models (for a review see, e.g., \cite{Branco:2011iw}),
or models with one doublet and one septet \cite{Kanemura:2013mc,Harris:2017ecz}.
As shown in \cite{Bento:2023weq},
in models with an 
SU(2)$_L\times $U(1)$_Y\times $U(1)$_{Y'}$ gauge group,
Eq.~\eqref{eq:cr_3y2} enforces instead a new tree-level parameter
\begin{equation}\label{eq:rho_prime}
    \rho^\prime \equiv \frac{\rho}{1 
    + \left( \displaystyle\frac{m^2_{Z^{\prime}}}{m_Z^2} - 1 \right) \sin^2 \alpha}
    = 1 \, ,
\end{equation}
where we have used Eq.~\eqref{eq:mz0_mz}.
The role of $\rho^\prime$ in an SU(2)$_L\times $U(1)$_Y\times $U(1)$_{Y'}$ gauge theory is
analogous to the role of $\rho$ in an SU(2)$_L\times $U(1)$_Y$ gauge theory. 

Using Eq.~\eqref{eq:def_sw},
it is convenient to replace \eq{treerho} with the following equivalent definition:
%
\begin{equation}\label{eq:true_def_rho}
    \rho \equiv
    \frac{2 G_F m_W^4}{m_Z^2 \left( 2 G_F m_W^2 - \sqrt{2} \pi \alpha\lsub{\rm EM} \right)} \,.
\end{equation}
In particular,  \eq{eq:def_sw} is a suitable definition in both SU(2)$_L\times $U(1)$_Y$ and
in SU(2)$_L\times $U(1)$_Y\times $U(1)$_{Y'}$ gauge theories, independently of how one chooses to define $c_W$.
%
Finally, using Eq.~\eqref{eq:rho_prime} we arrive at
\begin{equation}\label{eq:rho_m1}
    \rho - 1 = \left( r - 1 \right)
    \sin^2 \alpha \, ,
\end{equation}
where 
\beq \label{rdef}
r\equiv \frac{m_{Z^\prime}^2}{m_Z^2}\,,
\eeq
and the magnitude of $\sin \alpha$ is controlled by 
$\left( \mathcal{M}^2 \right)_{12}$.  In particular, 
\beq
\sin^2 2\alpha=\frac{4\bigl[(\mathcal{M}^2)_{12}\bigr]^2}{(m_Z^2-m_{Z'}^2)^2}\,.
\eeq
It is
useful to eliminate $\sin\alpha$ in favor of the parameter $r_{12}^2$
defined below:
\begin{equation} \label{r12def}
    r_{12}^2 \equiv \left(\frac{(\mathcal{M}^2)_{12}}{m_{Z^0}^2}\right)^2
    = \frac{(1-r)^2\sin^2\alpha \cos^2\alpha }
    {\left[1 - (1-r) \sin^2\alpha \right]^2} \,.
\end{equation}
Then,
\begin{equation}\label{eq:s_alpha}
    \sin^2\alpha =
    \frac{1 - r + 2 r_{12}^2 - \sqrt{(1-r)^2 - 4 r \, r_{12}^2}}
    {2(1-r)(1+r_{12}^2)} \, .
\end{equation}
Combining Eq.~\eqref{eq:rho_m1} with Eq.~\eqref{eq:s_alpha} yields,
\begin{equation}\label{eq:rho_m1_rs}
    \rho - 1 = \frac{-1 + r - 2 r_{12}^2 + \sqrt{(1-r)^2 - 4 r \, r_{12}^2}}
    {2(1+r_{12}^2)} \, ,
\end{equation}
which is a monotonically decreasing function of $r_{12}$.
This fact will be instrumental
in the results obtained in Section~\ref{sec:DMvsPrecision}.

In the dark matter models considered in Sections~\ref{sec:dark_matter} and \ref{sec4}, we shall assume that $0<m_{Z'}<m_Z$ or equivalently
$0<r<1$.  Then \eq{eq:rho_m1} implies that $\rho-1<0$.  It is then convenient to use \eq{eq:rho_m1_rs} to obtain $r_{12}^2$ as a function of $r$ and $\rho$:
\beq \label{r12solve}
r_{12}^2=\frac{(1-\rho)(\rho-r)}{\rho^2}\,.
\eeq

\section{Dark matter}
\label{sec:dark_matter}

A dark matter candidate is commonly included in the dark $Z^\prime$ model either by adding an
inert scalar field or by adding an SU(2)$_L\times$U(1)$_Y$ singlet Dirac fermion with a nonzero $\textrm{U}(1)_{Y^\prime}$
charge. In what follows, we shall consider such a Dirac fermion, denoted by~$\chi$,  as the dark matter
candidate.

Thus, the dark Lagrangian is given by
\begin{equation}
    \mathcal{L}_{\textrm{DM}} = i \overline{\chi} \slashed{D} \chi 
    - m_\chi \overline{\chi} \chi \, ,
\end{equation}
where the covariant derivative
can be expanded as
\begin{equation}\label{eq:conv_der_DM}
    D_\mu = \partial_\mu + i g_X Y^\prime \eta(s_\alpha Z _\mu+ c_\alpha \, Z^\prime_\mu) \, ,
\end{equation}
with $s_\alpha\equiv \sin\alpha$,
$c_\alpha \equiv \cos \alpha$,
and $\eta$ is defined in \eq{etadef}.
We note the difference between Eq.~\eqref{eq:conv_der_DM}
and the corresponding expressions given in
\cite{Bauer:2017qwy,Pospelov:2007mp}
\begin{equation}\label{eq:conv_der_DM_theirs}
    D_\mu \overset{\textrm{\cite{Bauer:2017qwy,Pospelov:2007mp}}}{\supset}
    \partial_\mu + i g_X Y^\prime  Z^\prime_\mu \, .
\end{equation}
The approximation in Eq.~\eqref{eq:conv_der_DM_theirs} requires
two assumptions: (i)  $\eta \approx 1$, and 
(ii) $c_\alpha \approx 1$.   Note that the latter is not guaranteed
to be true in all models even if $\epsilon = 0$. Indeed, by adding
non-inert, non-singlet scalars to the theory, $c_\alpha$ is controlled
by $\epsilon$ and $g_X$. Because $g_X$ still needs to be significant to explain
the correct abundance of DM,
we may constrain (and perhaps exclude) models using precision electroweak observables, such as 
the $\rho$ parameter
as defined in \eq{eq:true_def_rho}.

The interaction term of the $Z^\prime$ boson with a SM fermion $\psi$ is given by
\begin{equation}\label{eq:ffZd_corr}
    \mathcal{L}_{\textrm{int}} \supset - \frac{g}{2 c_W}
    \overline{\psi} \, \gamma^\mu \left( g_V - g_A \gamma_5 \right) \psi Z^\prime_\mu \, ,
\end{equation}
where
\begin{align}
    g_V &= \left( 2 Q s_W^2 - T_3 \right) s_\alpha + 
    \left( \eta t_W \epsilon \right) \left( 2Q - T_3 \right) c_\alpha \, , \label{eq:ffZd_corr_gv} \\[3pt]
    g_A &= - T_3 s_\alpha - \left( \eta t_W \epsilon \right) T_3 c_\alpha \,,\label{eq:ffZd_corr_ga}
\end{align}
and $Q=T_3+Y$.
In \cite{Bauer:2017qwy}, the authors employ 
\begin{equation}\label{eq:ffZd_bauer}
    \mathcal{L}_{\textrm{int}} 
    \overset{\textrm{\cite{Bauer:2017qwy}}}{\supset}
    \overline{\psi} \, \gamma^\mu \left( - \epsilon \, e \, Q \right) \psi Z^\prime_\mu \, ,
\end{equation}
in the small $\epsilon$
and small $m_{Z^\prime}$ approximation.
In particular, in the case of a scalar field that is neutral with respect to
SU(2)$_L\times$U(1)$_Y$ and has a nonzero U(1)$_{Y'}$ charge
(treated in \cite{Bento:2023weq}), one obtains $s_\alpha\simeq -
\eta t_W\epsilon$ and $c_\alpha\simeq 1$ after dropping terms of order $\mathcal{O}(\epsilon^2)$
and $\mathcal{O}(r)$.  Inserting these results into \eqst{eq:ffZd_corr}{eq:ffZd_corr_ga} reproduces
\eq{eq:ffZd_bauer}.

In the following, we assume that the DM candidate $\chi$ is in thermal equilibrium in the early Universe.   The velocity averaged cross section for $\overline{\chi}\chi$ annihilation is given by  $\expval{\sigma_{\chi \chi} v}\simeq 2\times 10^{-26}~{\rm cm}^3\, {\rm s}^{-1}\simeq 1.7\times 10^{-9}~{\rm GeV}^{-2}$ (the latter in natural units) for values of
$m_\chi\gsim 10$~GeV~\cite{Planck:2018vyg}, under the assumption that $\chi$ particles saturate the observed DM abundance today.
As the Universe evolves and the temperature drops, a point is reached where the DM decouples from the thermal bath and it freezes out~\cite{Profumo:2017hqp,Bauer:2017qwy}.
Freeze-out in the U(1)$_{Y^\prime}$ model considered above may be accessed through two main regimes
which we now briefly consider---the
characteristic and secluded regimes.

\subsection{Freeze-out: characteristic regime}

The characteristic regime corresponds to
the mass ordering given by $m_{Z^\prime} > m_\chi > m_e$, where the
dominant annihilation mechanism is the $s$-channel scattering process
$\overline{\chi} \chi \rightarrow {Z^\prime}^*  \rightarrow \bar{f} \bar{f}$.
Then, the velocity averaged annihilation cross section is given by~\cite{Bauer:2017qwy}
\begin{equation}\label{eq:bauers_t_avg_cs}
    \expval{\sigma_{\chi \chi} v} \overset{\textrm{\cite{Bauer:2017qwy}}}{\approx}
    \frac{m_\chi^2}{\pi m_{Z^\prime}^4} \left( \epsilon e g_X Y^\prime \right)^2 \, ,
\end{equation}
under the assumption that $m_\chi \gg m_e$ and $m_{Z^\prime} \gg m_\chi$.
By assuming $Y^\prime = 1$ (and correcting some minor misprints in \cite{Bauer:2017qwy}), 
we obtain the correct DM abundance with
\begin{equation}\label{eq:bauers_char_abund}
    \frac{1.7 \times 10^{-9}}{\textrm{GeV}^{2}}
    \overset{\textrm{\cite{Bauer:2017qwy}}}{\approx}
    \frac{0.038}{\textrm{GeV}^{2}}
    \left( \frac{m_\chi}{0.01 \, \textrm{GeV}} \right)^2
    \left( \frac{0.1 \, \textrm{GeV}}{m_{Z^\prime}} \right)^4
    (\epsilon \, g_X)^2 \, ,
\end{equation}
which, after fixing the masses, yields a value for $\epsilon \, g_X$
in agreement with \cite{Izaguirre:2014dua,Davoudiasl:2015hxa}.
However, as there are contributions from both vector and axial interactions,
a full model computation of Eq.~\eqref{eq:bauers_t_avg_cs} is needed,
which is provided in Appendix~\ref{app}, to obtain a more precise result.
When \eq{eq:bauers_t_avg_cs} is replaced by the expression obtained in
\eq{appeq}, we find that the numerical result obtained in \eq{eq:bauers_char_abund}
is modified by a factor of order unity.

For models with a light $Z'$ where $m_\chi<m_{Z'}<10$~GeV, the characteristic regime is ruled out
by the bound obtained in Fig.~46 of \cite{Planck:2018vyg}, which is derived from the
cosmic microwave background (CMB) data.

\subsection{Freeze-out: secluded regime}

The secluded regime, which was advocated in \cite{Pospelov:2007mp} and explored further in \cite{Evans:2017kti}, corresponds to 
the mass ordering $m_\chi > m_{Z^\prime} > m_e$, where the
dominant annihilation mechanism is $\overline{\chi} \chi  \rightarrow Z^\prime Z^\prime$ via $t$-channel $\chi$-exchange.
In contrast to the characteristic regime, the CMB bound cited above does not rule out a light $Z'$ for $m_\chi>10$~GeV.
The corresponding velocity averaged annihilation cross section is given by \cite{Bauer:2017qwy}
\begin{equation}\label{eq:aux1}
    \expval{\sigma_{\chi \chi} v} 
    \overset{\textrm{\cite{Bauer:2017qwy}}}{\approx}
    \frac{g_X^4 {Y^\prime}^4}{8 \pi m_\chi^2} \, ,
\end{equation}
under the assumption that $m_{\chi} \gg m_{Z^\prime}$.  Assuming again that $Y'=1$, we obtain the correct DM abundance
with
\begin{equation}\label{eq:bauer_secl_abun}
    1.7 \times 10^{-9} \, \textrm{GeV}^{-2}
    \overset{\textrm{\cite{Bauer:2017qwy}}}{\approx} 0.04 \,
    \frac{g_X^4 }{m_\chi^2} \, ,
\end{equation}
which, after fixing the mass $m_\chi$, yields a value for $g_X$.

A more precise analysis that employs the full SU(2)$_L\times $U(1)$_Y\times $U(1)$_{Y'}$ model
will modify the results obtained above.
Using Eq.~\eqref{eq:conv_der_DM}, the results of Eqs.~\eqref{eq:aux1}--\eqref{eq:bauer_secl_abun}
are modified as follows:
\beqa
    \expval{\sigma_{\chi \chi} v} 
   & \approx &
    \frac{g_X^4 \, \eta^4 \, c_\alpha^4 \, {Y^\prime}^4}{8 \pi m_\chi^2} \, ,\label{eq:ours_secl_thermal_cs} \\
    1.7 \times 10^{-9} \, \textrm{GeV}^{-2}
   &  \approx &  0.04 \,
    \frac{g_X^4 \, \eta^4 \, c_\alpha^4}{m_\chi^2} \, .\label{eq:ours_secl_abun}
\eeqa
After fixing $m_\chi$ and $m_{Z'}$, we may constrain the values of $g_X$ and $\epsilon$
[after employing Eq.~\eqref{eq:s_alpha}, which determines the value of $c_\alpha$].
In particular, even if $\epsilon \ll 1$, 
one cannot assume in general that $c_\alpha\simeq 1$.

\section{\label{sec:DMvsPrecision}Dark matter and the electroweak $\rho$ parameter}
\label{sec4}

By considering $Z^\prime$ as the vector portal to a dark
sector, we may infer a relation between the dark matter relic
density and the parameters $g_X$ and $\epsilon$.
In Section~\ref{sec:dark_matter} we saw that by fixing
$m_\chi$ and $m_{Z^\prime}$ we could constrain 
$g_X$ and~$\epsilon$ through Eq.~\eqref{eq:bauers_char_abund}
and Eq.~\eqref{eq:ours_secl_abun}. The secluded
regime is largely independent of $\epsilon$,
and we will use this regime to provide the following instructive example.

Consider an SU(2)$_L\times $U(1)$_Y\times $U(1)$_{Y'}$ model
that possesses scalar multiplets $\varphi_i$ beyond the SM Higgs doublet that are non-inert (i.e., $v_i\neq 0$)
and charged under both U(1)$_Y$ and U(1)$_{Y'}$.  
In particular, we assume a parameter regime
where $\epsilon\ll 1$ and $r=m^2_{Z'}/m_Z^2\ll 1$.
 Using \eq{eq:s_alpha},
\beq
c_\alpha^2=\frac{1}{1+r_{12}^2}+\mathcal{O}(r)\,.
\eeq
\Eq{eq:mzzprime}
then yields $r_{12}^2=\bigl[(\mathcal{M}^2)_{12}\bigr]^2/m_{Z^0}^4\sim g_X^2/g^2$,
under the assumption that\footnote{Note that the assumption that \eq{order1} is satisfied can be consistent with the requirement that there exists a SM-like Higgs boson $h$ in the scalar spectrum (as indicated by the LHC Higgs data~\cite{ATLAS:2022vkf,CMS:2022dwd}) if the decoupling limit is 
realized~\cite{Gunion:2002zf} (where all new scalars beyond the SM are significantly heavier than $h$).} 
\beq \label{order1}
\frac{4c_W}{v^2}\sum_iy_i y^\prime_i v_i^2\sim\mathcal{O}(1)\,.
\eeq
For example, if $m_\chi = 20 \, \mathrm{GeV}$ then Eq.~\eqref{eq:ours_secl_abun} yields
\begin{equation}\label{eq:epsion_gx}
 g_X \sim 0.0645\,.
\end{equation}

\begin{figure}[t!]
\includegraphics[width=8.75cm]{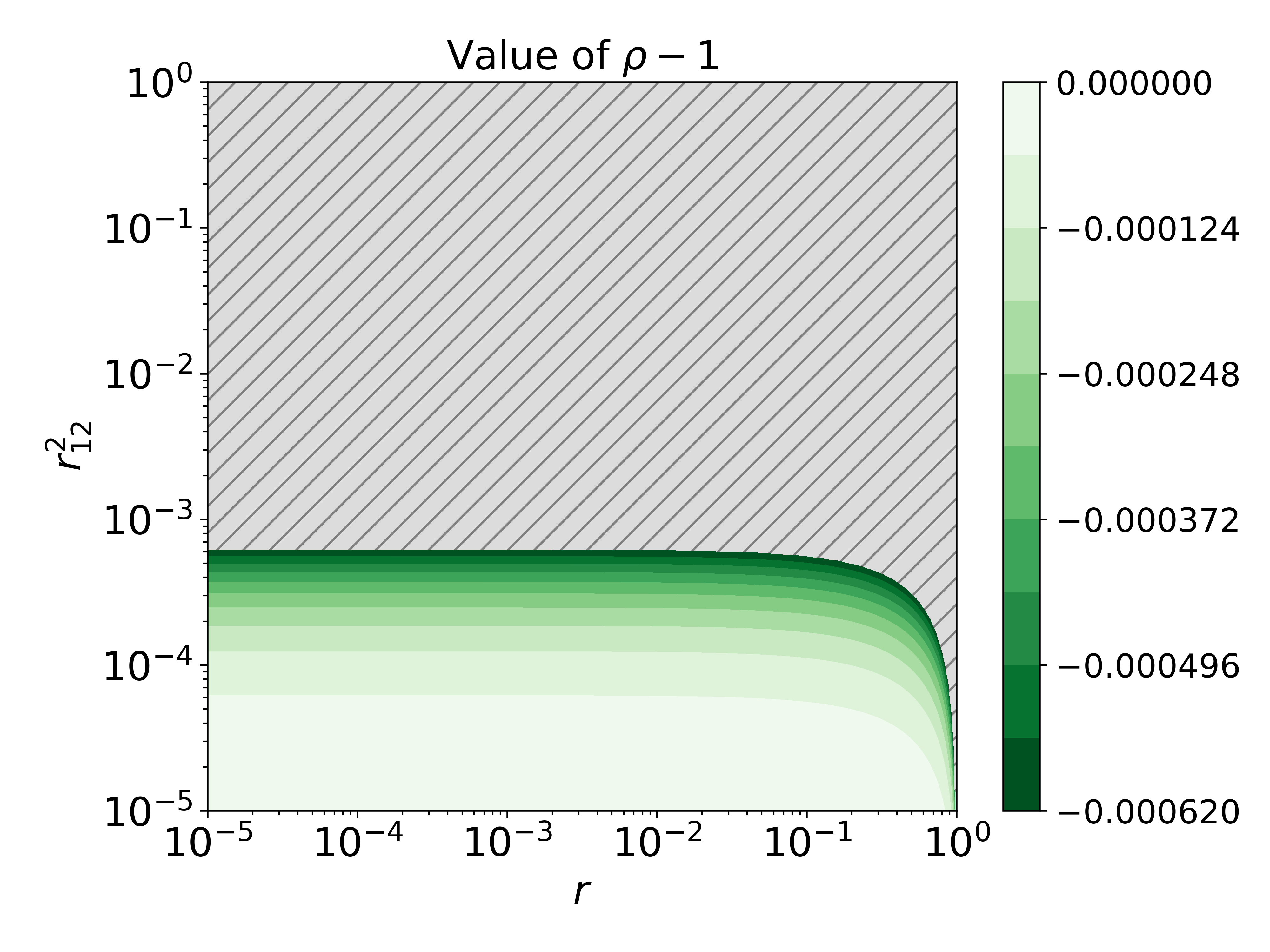}
\centering
\caption{Variation of $\rho - 1$ with $r_{12}^2$ and $r$ obtained from \eq{r12solve}.  
Note that $\rho-1<0$ in light of \eq{eq:rho_m1}.  Values of $\rho-1$ that are less than [more than] $5\sigma$ below the central
value given in \eq{rhopdg} (obtained in the global electroweak fit of \cite{EWPDG}) are exhibited by the shaded green [gray
hatched] regions of the plot.\label{fig:rhom1}}
\end{figure}

As $g_X$ becomes larger, so does 
$r_{12}^2$. Because Eq.~\eqref{eq:rho_m1_rs} was
determined to be a monotonically decreasing function with $r_{12}^2$, the quantity
$\rho - 1$ gets more negative
with larger $r_{12}$. Thus, a large
$g_X$ pushes towards a larger negative value of $\rho - 1$.
An illustration of this effect is exhibited in Fig.~\ref{fig:rhom1}.
It can be seen that for $r \ll 1$, the contribution of $r$
to $\rho - 1$ is small. Then, we may approximate Eq.~\eqref{eq:rho_m1_rs} by
\begin{equation}
    \rho - 1 = - \frac{r_{12}^2}{1 + r_{12}^2} + \mathcal{O}(r) \, .
\end{equation}

Using $r_{12}^2\sim g_X^2/g^2\sim 2.34 g_X^2$, we end up with
\begin{equation}
    \rho - 1 \sim  - 0.0096\,.
\end{equation}
This value, which lies in the gray hatched region of Fig.~\ref{fig:rhom1}, is inconsistent with the global electroweak fit value of~\cite{EWPDG}
\beq \label{rhopdg}
\rho_0=1.00038\pm 0.00020\,.
\eeq
Note that as defined by the authors \pagebreak
in \cite{EWPDG}, $\rho_0=1$ exactly in the SM, and a deviation from $\rho_0=1$ can be interpreted as a consequence of new physics beyond the SM (under the assumption that it is a small perturbation that does not significantly affect other electroweak radiative corrections).  
In the present context, we can assume that the deviation from $\rho_0=1$ is due primarily to the tree-level effect exhibited in \eq{eq:rho_m1_rs}.

As a second example, consider an extended Higgs sector that contains
an SU(2)$_L\times$U(1)$_Y$ singlet scalar~$\varphi$ with a U(1)$_{Y'}$ charge of $y'=1$.
\Eqs{em12}{r12def} yield:
\beq \label{r12approx}
r_{12}^2=\eta^2 t_W^2 \epsilon^2\,.
\eeq
Under the assumption that $|\epsilon|\ll 1$ and $1-r\sim\mathcal{O}(1)$, 
we can employ \eq{eq:rho_m1_rs}
to obtain:
\beq \label{next}
\rho-1= \frac{r_{12}^2}{r-1} +\mathcal{O}(r_{12}^4)\,.
\eeq
After inserting \eq{r12approx} into the above result and noting that $\eta=1+\mathcal{O}(\epsilon^2)$ [cf.~\eq{etadef}], we end up with
\beq \label{rhodark}
\rho-1=-\,\frac{\epsilon^2 t_W^2}{1-r}+\mathcal{O}(\epsilon^4)\,,
\eeq
in agreement with a result previously obtained in \cite{Bento:2023weq}.
For example, assuming that the true value of $\rho_0$ is no more than $5\sigma$ below the central value given in \eq{rhopdg}, 
one can deduce an upper limit of $|\epsilon|\lsim 0.046$.
This is one of a number of experimental observables that can be used to constrain the value of $\epsilon$ (e.g., see Fig.~6 of \cite{Evans:2017kti}).
 
Finally, we remark that in contrast to the dark matter models considered in Section~\ref{sec:dark_matter}, in models of asymmetric dark matter, the
cross section required to annihilate the symmetric component of the dark matter must be roughly $2$--$3$ times larger than the corresponding
annihilation cross sections of symmetric thermal dark matter models~\cite{Graesser:2011wi,Iminniyaz:2011yp}.   The more efficient annihilation cross section required by 
asymmetric dark matter models implies a larger value of $g_X$, which would yield an even smaller allowable parameter space in light of
\eq{eq:ours_secl_abun}.

\section{\label{sec:Concl}Discussion and Conclusions}

In gauge theories with an SU(2)$_L\times $U(1)$_Y\times $U(1)$_{Y'}$ gauge group,
we have examined models whose low energy behavior appears to be equivalent to
a model of a photon that mixes with a new light neutral gauge boson (e.g., the dark
photon).  In the literature, the latter is often employed in models that propose to explain
the observed DM relic density, without considering the impact of the full 
 SU(2)$_L\times $U(1)$_Y\times $U(1)$_{Y'}$ model on $Z$ physics observables.
It is remarkable that the implications of the $\rho$ parameter 
alone considered in Section~\ref{sec4} provide a simple illustration that the
two requirements---the origin of the DM and the consistency with $Z$ physics observables---can
easily be at odds with each other.

Additional constraints will arise 
when the impact of the SU(2)$_L\times $U(1)$_Y\times $U(1)$_{Y'}$ gauge theory on
other precision electroweak observables is considered.
For example, three tree-level
shifts to the SM electroweak Lagrangian due to the effect of the mixing of the interaction eigenstates $Z^0$ and $X$ are identified in \cite{Davoudiasl:2023cnc},
which are denoted by $\Delta_1$, $\Delta_2$, and $\Delta_3$.  In particular, $\Delta_1$ is related to the tree-level $\rho$ parameter
via $\rho=(1+\Delta_1)^{-1}$.  The other two quantities $\Delta_2$ and $\Delta_3$ correspond to shifts of the SM neutral
current and the electromagnetic current to the physical $Z$ boson, respectively.   Indeed, there have been a number
of authors (e.g., see \cite{Davoudiasl:2012ag,Davoudiasl:2012qa,Curtin:2014cca,Davoudiasl:2015bua,Harigaya:2023uhg,Davoudiasl:2023cnc,Hook:2010tw,Sun:2023kfu})
who have performed fits to the precision electroweak data as a way of deducing interesting constraints
on dark photon (dark $Z$) models.
It would be of interest to generalize such studies to examine the full impact of the SU(2)$_L\times $U(1)$_Y\times $U(1)$_{Y'}$ gauge 
theory.  One such approach is to employ an
effective field theory technique that includes all the gauge boson fields of the
SU(2)$_L\times $U(1)$_Y\times $U(1)$_{Y'}$ model (e.g., see \cite{Aebischer:2022wnl}).   We shall explore
the implications of a more complete analysis in a future publication.

\section*{Acknowledgements}

We are grateful for discussions with Hooman Davoudiasl and Stefania Gori.
The work of M.P.B. is supported in part by the Portuguese
Funda\c{c}\~{a}o para a Ci\^{e}ncia e Tecnologia\/ (FCT)
under contract SFRH/BD/146718/2019.
The work of M.P.B. and J.P.S. is supported in part by FCT under Contracts
CERN/FIS-PAR/0008/2019,
PTDC/FIS-PAR/29436/2017,
UIDB/00777/2020,
and UIDP/00777/2020;
these projects are partially funded through POCTI (FEDER),
COMPETE,
QREN,
and the EU.
The work of H.E.H. is supported in part by the U.S. Department of Energy Grant No. DE-SC0010107.
H.E.H. is grateful for the hospitality and support during his visit to the Instituto Superior
T\'{e}cnico, Universidade de Lisboa, where their work was initiated.

\appendix

\section{The annihilation cross section in the characteristic regime revisited}
\label{app}

In \cite{Bauer:2017qwy}, the authors derive the cross section
for dark matter annihilation in the characteristic regime.
There is a misprint in the equation that when corrected takes the form
\begin{equation}\label{eq:bauers_cross}
    \sigma_{\chi \chi} \overset{\textrm{\cite{Bauer:2017qwy}}}{=}
    \frac{\left( \epsilon e g_X Y^\prime \right)^2}{12 \pi s} 
    \frac{ \left( s + 2 m_\chi^2 \right) \left( s + 2 m_e^2 \right)}
    { (s - m_{Z^\prime}^2)^2 + m_{Z^\prime}^2 \Gamma_{Z^\prime}^2} \frac{\beta_f}{\beta_i} \, ,
\end{equation}
where
\begin{equation}
    \beta_i = \sqrt{1 - \frac{4 m_\chi^2}{s}} 
    \quad \mathrm{and} \quad
    \beta_f = \sqrt{1 - \frac{4 m_e^2}{s}} \, .
\end{equation}

With the relative velocity $v$ defined
as $v = 2 \beta_i$, we may compute the
velocity averaged annihilation cross section using Eq.~\eqref{eq:bauers_cross},
\begin{equation}\label{eq:bauers_thermally_avg_cs}
    \expval{\sigma_{\chi \chi} v} \overset{\textrm{\cite{Bauer:2017qwy}}}{=}
    \frac{\left( \epsilon e g_X Y^\prime \right)^2}{2 \pi} 
    \frac{\sqrt{m_\chi^2 - m_e^2} \left( 2 m_\chi^2 + m_e^2 \right)}
    {m_\chi \left( m_{Z^\prime}^2 - 4 m_\chi^2 \right)^2} \, ,
\end{equation}
where we have assumed $\Gamma_{Z^\prime} / m_{Z^\prime} \ll 1$.
Eq.~\eqref{eq:bauers_thermally_avg_cs} can be further simplified
under the assumption that $m_\chi \gg m_e$ and $m_{Z^\prime} \gg m_\chi$, which yields 
the approximate result,
\begin{equation}\label{app:eq:bauers_thermally_avg_cs_approx}
    \expval{\sigma_{\chi \chi} v} \overset{\textrm{\cite{Bauer:2017qwy}}}{\approx}
    \frac{m_\chi^2}{\pi m_{Z^\prime}^4}
    \left( \epsilon e g_X Y^\prime \right)^2 \, ,
\end{equation}
in agreement with Eq.~\eqref{eq:bauers_t_avg_cs}.

In the full SU(2)$_L\times $U(1)$_Y\times $U(1)$_{Y'}$ model, we obtain the following expressions for the annihilation cross
section and corresponding velocity averaged cross section:
\begin{align}
    \sigma_{\chi \chi} =&
    \frac{1}{12 \pi s} \left( \frac{g \, g_X}{2 c_W} \eta c_\alpha Y^\prime \right)^2
    \frac{s + 2 m_\chi^2}
    {(s - m_{Z^\prime}^2)^2 + m_{Z^\prime}^2 \Gamma_{Z^\prime}^2} 
    \times \nonumber \\
    &\left[ g_V^2 (s + 2 m_e^2) + g_A^2 (s - 4 m_e^2) \right] \frac{\beta_f}{\beta_i}
     \, ,
\end{align}
and
\begin{align}
    \expval{\sigma_{\chi \chi} v} =&
    \frac{1}{2 \pi} \left( \frac{g \, g_X}{2 c_W} \eta c_\alpha Y^\prime \right)^2
    \times \nonumber \\
    &\frac{\sqrt{m_\chi^2 - m_e^2} \left[ 2 m_\chi^2 (g_V^2 + g_A^2) + m_e^2 (g_V^2 - 2 g_A^2) \right]}
    {m_\chi \left( m_{Z^\prime}^2 - 4 m_\chi^2 \right)^2} 
    \,,
\end{align}
in agreement with Eq.~(4.1) of \cite{Alves:2015pea} in the absence of an axial vector coupling of the DM to the $Z'$.
By using the same set of approximations, $m_\chi \gg m_e$ and $m_{Z^\prime} \gg m_\chi$,
employed in the derivation of Eq.~\eqref{app:eq:bauers_thermally_avg_cs_approx},
we obtain the thermally averaged annihilation cross section, \\
\beqa 
&& \phantom{line}\label{appeq} \\[-33pt]
&&    \expval{\sigma_{\chi \chi} v} \approx 
    \frac{ m_\chi^2}{\pi m_{Z^\prime}^4} (g_V^2 + g_A^2)
    \left( \frac{g \, g_X}{2 c_W} \eta c_\alpha Y^\prime \right)^2 \,,\nonumber 
\eeqa
which yields an $\mathcal{O}(1)$ correction to the result quoted in \eq{eq:bauers_t_avg_cs}.

\bibliographystyle{elsarticle-num} 
\bibliography{biblio}

\end{document}